\documentclass[twocolumn,prb,showpacs]{revtex4}

\usepackage{graphicx}
\usepackage{dcolumn}
\usepackage{bm}
\usepackage{amssymb, amsmath}
\usepackage{color}
\usepackage{slashed} 

\begin{document}
\title{Electromagnetic properties of impure superconductors with
  pair-breaking processes}

\author{Franti\v{s}ek Herman and Richard Hlubina}

\affiliation{Department of Experimental Physics, Comenius University,
  Mlynsk\'{a} Dolina F2, 842 48 Bratislava, Slovakia}

\begin{abstract}
Recently a generic model has been proposed for the single-particle
properties of gapless superconductors with simultaneously present
pair-conserving and pair-breaking impurity scattering (the so-called
Dynes superconductors). Here we calculate the optical conductivity of
the Dynes superconductors.  Our approach is applicable for all
disorder strengths from the clean up to the dirty limit and for all
relative ratios of the two types of scattering, nevertheless the
complexity of our description is equivalent to that of the widely used
Mattis-Bardeen theory.  We identify two optical fingerprints of the
Dynes superconductors: (i) the presence of two absorption edges and
(ii) a finite absorption at vanishing frequencies even at the lowest
temperatures.  We demonstrate that the recent anomalous optical data
on thin MoN films can be reasonably fitted by our theory.
\end{abstract}
\maketitle

\section{Introduction}
In the limit of low temperatures and low excitation energies, the
single-particle properties as well as the two-particle response
functions of metals are governed by elastic scattering on impurities.
Provided that the effect of impurities can be described by classical
fields, the normal-state properties are largely independent of the
precise nature of those fields. However, once the metal becomes
superconducting, a sharp distinction can be made between two types of
impurity scattering: If the impurity potential is time
reversal-invariant, the corresponding scattering does not destroy the
pairing, and Anderson gave a very general argument that neither the
thermodynamic properties, nor the tunneling density of states of the
superconductor can change with respect to the clean
case.\cite{Anderson59} On the other hand, Abrikosov and Gor'kov have
shown that, if the impurity potential does not respect time-reversal
symmetry, then the corresponding scattering becomes pair breaking,
thereby affecting both, the thermodynamics and the tunneling density
of states of the superconductor.\cite{Abrikosov61}

Surprisingly, experimentally it has been found that, quite often, the
tunneling density of states of dirty superconductors $N(\omega)$ is
described neither by the Anderson result, nor by the Abrikosov-Gor'kov
theory, but rather by the following simple phenomenological Dynes
formula:\cite{Dynes78,Noat13,Szabo16}
\begin{equation}
N(\omega)=N_0{\rm Re}\left[\frac{\omega+i\Gamma}
{\sqrt{(\omega+i\Gamma)^2-{\overline{\Delta}}^2}}\right],
\label{eq:dynes}
\end{equation}
where $N_0$ is the normal-state density of states, $\overline{\Delta}$
is the ideal gap of the dirty superconductor, and $\Gamma$ describes
its filling by in-gap states.  The square root in Eq.~(\ref{eq:dynes})
has to be taken so that its imaginary part is positive and we keep
this convention throughout this paper. The microscopic origin of the
Dynes formula had been unclear for a long time, but in a recent
paper\cite{Herman16} we have shown that the tunneling density of
states described by Eq.~(\ref{eq:dynes}) is realized in systems with
pair-breaking classical disorder, provided that the pair-breaking
potentials have a Lorentzian distribution with width $\Gamma$; we did
not need to make any assumptions about the nature of the
pair-conserving disorder.

More importantly, we have also found that the full $2\times 2$
Nambu-Gor'kov Green's function of a Dynes superconductor with density
of states Eq.~(\ref{eq:dynes}) can be described by just three
parameters: the ideal gap $\overline{\Delta}$ and the pair-conserving
and pair-breaking scattering rates $\Gamma_s$ and $\Gamma$,
respectively. The final result for the Green's function can be written
in the following elegant way\cite{Herman17}
\begin{equation}
\hat{G}({\bf k},\omega)= \frac{1}{2}\slashed{\partial}
\ln \left[\varepsilon_{\bf k}^2-\epsilon(\omega)^2\right],
\label{eq:dynes_green}
\end{equation}
where $\slashed{\partial} = \tau_0\frac{\partial}{\partial\omega}
-\tau_1\frac{\partial}{\partial\overline{\Delta}}
-\tau_3\frac{\partial}{\partial\varepsilon_{\bf k}^{}}$ and $\tau_i$
are the Pauli matrices. In Eq.~(\ref{eq:dynes_green}) we have also
introduced the function
\begin{equation}
\epsilon(\omega)
=\sqrt{(\omega+i\Gamma)^2-{\overline\Delta}^2}+i\Gamma_s.
\label{eq:epsilon}
\end{equation}
Note that $\epsilon(\omega)$ of a Dynes superconductor differs from
its value in a clean superconductor in a minimalistic way:
pair-conserving processes are taken into account by the replacement
$\epsilon(\omega)\rightarrow\epsilon(\omega)+i\Gamma_s$, whereas
pair-breaking processes are described by the frequency shift
$\omega\rightarrow\omega+i\Gamma$ implicit in Eq.~(\ref{eq:dynes}).

In Ref.~\onlinecite{Herman17} we have pointed out that the Green
function Eq.~\eqref{eq:dynes_green} is analytic in the upper
half-plane and has the correct large-frequency asymptotics; therefore
it satisfies the known exact sum rules.  We have also proven that the
diagonal components of the corresponding spectral function are
positive-definite, as it should be. Moreover, in the three limiting
cases of either $\Gamma=0$, or $\Gamma_s=0$, or $\overline{\Delta}=0$,
Eq.~\eqref{eq:dynes_green} reproduces the well-known results. In
particular, for $\overline{\Delta}=0$ Eq.~\eqref{eq:dynes} describes a
normal metal with the total scattering rate
\begin{equation}
\Gamma_n=\Gamma_s+\Gamma.
\label{eq:gamma_n}
\end{equation}
Therefore, although the Green function Eq.~\eqref{eq:dynes_green}
has been derived only for a special distribution of pair-breaking
fields within the coherent potential approximation, we believe that it
represents a generic Green function of the Dynes superconductors.

In order to support this point of view, in Ref.~\onlinecite{Herman17}
we have compared the results of high-resolution angle resolved
photoemission spectroscopy to the Dynes phenomenology, and we have
found that Eq.~\eqref{eq:dynes_green} can fit the spectra of the
cuprates\cite{Kondo15} with reasonable accuracy.  This success has
motivated us to investigate also the two-particle properties of the
Dynes superconductors, and in this paper we will discuss the arguably
most important of such properties, namely the optical conductivity
$\sigma(\omega)$.

The specific problem which we will be interested in is the following.
In the normal state, the optical properties are governed only by the
total scattering rate $\Gamma_n$, but in the superconducting state the
partition of $\Gamma_n$ into its components $\Gamma_s$ and $\Gamma$
must obviously be of crucial importance, and different partitionings
will lead to different functions $\sigma(\omega)$. Our goal in this
paper will be to describe qualitative effects of such partitionings.

Electrodynamics of superconductors has been studied since the early
days of the BCS theory, starting with the classic paper by Mattis and
Bardeen which focused on the dirty limit with no pair-breaking
scatterings.\cite{Mattis58} In an important paper that appeared
relatively soon thereafter, Nam has developed a Green's function
approach to the problem,\cite{Nam67a} which stimulated many subsequent
works.\cite{Nam67b,Carbotte05,Coumou13,Zemlicka15} In a later
publication, Nam himself used this approach in studying the effect of
pair-breaking processes on $\sigma(\omega)$,\cite{Nam67b} but he
worked within the Abrikosov-Gor'kov theory which makes use of the Born
approximation and is not directly relevant to the Dynes
superconductors.

Basically the same approach as in Ref.~\onlinecite{Nam67b} has been
used in most of the published literature on the effects of pair
breaking.  As an example let us mention the recent combined tunneling
and microwave study of TiN films:\cite{Coumou13} the data has been
analyzed similarly as in Ref.~\onlinecite{Nam67b} and the authors did
find good agreement between theory and experiment in the low-disorder
limit where the Born approximation might be expected to work. However,
at the same time they have found strong discrepancy between the
Abrikosov-Gor'kov theory and experiment in the highly disordered films.
Yet another recent paper, technically equivalent to
Ref.~\onlinecite{Nam67b}, is devoted to the study of the influence of
pair breaking on the London penetration depth.\cite{Kogan13}

It should be mentioned that two important recent
papers\cite{Fominov11,Kharitonov12} treat electrodynamics of
superconductors with magnetic impurities going beyond the Born
approximation for magnetic scattering. Single-particle properties in
these papers are described on the level of the Marchetti-Simons
generalization of the Usadel equations,\cite{Marchetti02} which
reproduces Shiba's T-matrix result for the density of
states.\cite{Shiba68} Unfortunately, although scattering on a single
magnetic impurity is treated exactly within this approach,
Refs.~\onlinecite{Fominov11,Kharitonov12} can not be quantitatively
correct in case of dense magnetic impurities, since they do not take
multiple scattering on different impurities into account. This means
then that Refs.~\onlinecite{Fominov11,Kharitonov12} are applicable
only to superconductors with not too large density of magnetic
impurities, and therefore they can not describe electrodynamics of the
Dynes superconductors with a dense and broad distribution of
pair-breaking fields.\cite{Herman16}

To the best of our knowledge, the Dynes phenomenology has been taken
seriously only in the very recent paper Ref.~\onlinecite{Zemlicka15},
where Nam's dirty-limit formula for $\sigma(\omega)$ was evaluated
assuming that the tunneling density of states is given by
Eq.~\eqref{eq:dynes}. However, since neither the gap function
$\Delta(\omega)$, nor the full Green's function of a Dynes
superconductor were known at that time, the authors of
Ref.~\onlinecite{Zemlicka15} could only speculate about the correct
formula for $\sigma(\omega)$ in the dirty limit.  Moreover, it was
unclear how to obtain results for the Dynes superconductors away from
the dirty limit.  Both of these points will be addressed in this
paper.

The outline of this paper is as follows. In Section~2 we briefly
summarize Nam's results for the optical conductivity of a general
Eliashberg superconductor, making use of a slightly modified notation
with respect to the original paper Ref.~\onlinecite{Nam67a}.  In
Section~3 we apply Nam's analysis to the case of the Dynes
superconductors described by Eq.~\eqref{eq:dynes_green}. Our
presentation concentrates on various qualitative aspects of the
optical conductivity for all disorder strengths from the clean up to
the dirty limit and for all relative ratios $\Gamma/\Gamma_s$. In
Section~4 we analyze the recent anomalous optical data on thin MoN
films\cite{Simmendinger16} and we demonstrate that they can be
reasonably fitted by the theory for the Dynes superconductors.

\section{Optical conductivity: general theory}
We will assume that the Fermi surface is isotropic and the
normal-state spectrum is quadratic with effective mass $m$. Under
these assumptions also the optical conductivity
$\sigma(\omega)=\sigma^\prime(\omega)+i\sigma^{\prime\prime}(\omega)$
is isotropic and, within linear response theory, it can be written
as\cite{Marsiglio08,Rickayzen84,Nam67a}
\begin{equation}
\sigma(\omega)=\frac{i}{\omega+i 0^{+}}K(\omega),
\label{eq:kubo}
\end{equation}
where the current-current correlation function $K(\omega)$ on the
imaginary (Matsubara) axis reads, neglecting the vertex corrections,
\begin{eqnarray}
K(\omega_m)&=&D_0+\frac{e^2 v_F^2}{3}
\int \frac{d^3{\bf k}}{(2\pi)^3}\times
\nonumber 
\\
&&T\sum_{\omega_l}
{\rm Tr}
\left[{\hat G}(\mathbf{k},\omega_l + \omega_m)
{\hat G}(\mathbf{k},\omega_l)\right].
\label{eq:responsefunction}
\end{eqnarray}
The first term corresponds to the constant diamagnetic contribution
$D_0=ne^2/m$ which depends only on the density of the electrons $n$,
their mass $m$, and their charge $e$. The second term corresponds to
the paramagnetic contribution which is affected by superconductivity
as well as by the impurities. The momentum integration is taken over
the entire Brillouin zone, $T$ represents the temperature, and $v_F$
is the Fermi velocity.

\begin{figure*}[t!]
\includegraphics[width = 16 cm]{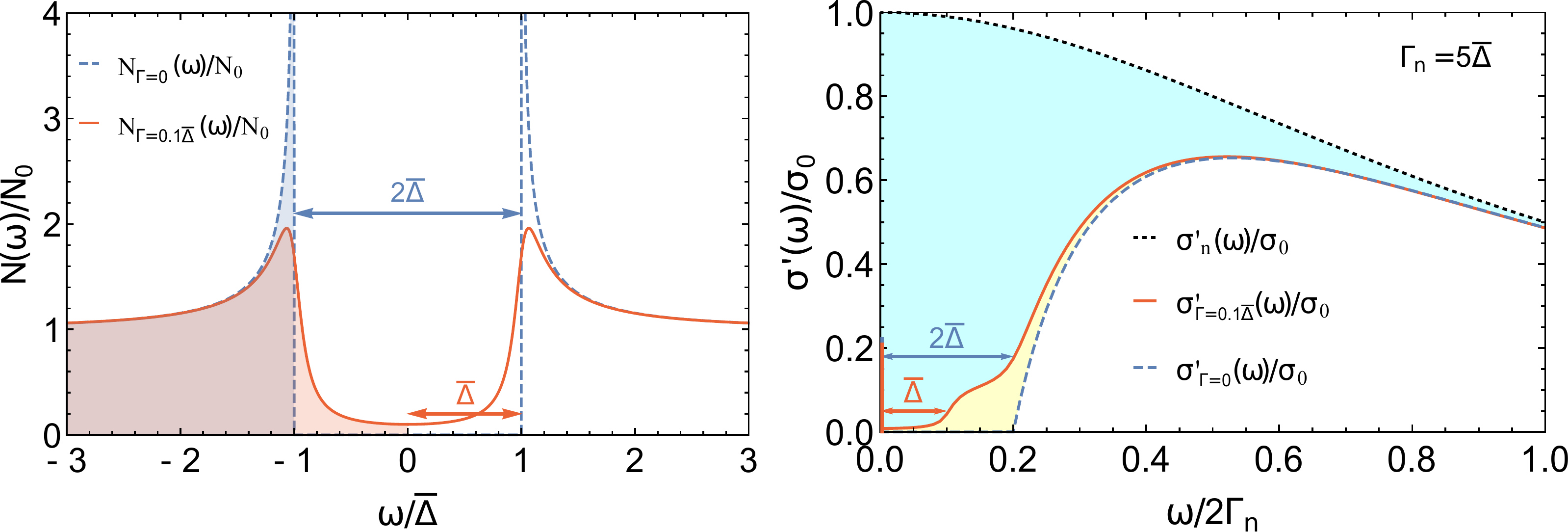}
\caption{Comparison, at $T=0$, of a normal metal with two
  superconductors: a superconductor without pair-breaking processes
  ($\Gamma=0$) and a Dynes superconductor
  ($\Gamma=0.1\overline{\Delta}$). Left panel: density of states for
  both superconductors, normalized with respect to the constant
  normal-state density $N_0$.  Shaded areas denote the occupied
  states. Right panel: real part of the optical conductivity
  $\sigma'(\omega)$ for the normal metal and for both
  superconductors. The total scattering rate is
  $\Gamma_n=5\overline{\Delta}$ in all cases. Shaded areas show the
  weight which is missing in order to satisfy the sum rule
  Eq.~\eqref{eq:cond_sum_rule}.  The missing weight is carried by the
  singular contribution to Eq.~\eqref{eq:conductivity}; this
  contribution is schematically shown as a bar at $\omega = 0$.}
\label{fig:Pedagog}
\end{figure*}

A note about the vertex corrections is in place here. It is well known
that, in general, in order to satisfy the Ward identities the vertex
corrections should be present. However, in our case they do not
appear, and the reasons are the same as in
Refs.~\onlinecite{Nam67a,Fominov11,Kharitonov12}. First, in order to
avoid the phase modulation of the superconducting order
parameter,\cite{Ambegaokar61} we work in the transverse gauge. Second,
since the impurity potential considered in Ref.~\onlinecite{Herman16}
is point-like, also the normal-state vertex corrections to
conductivity vanish in the CPA formalism.\cite{Velicky69,Elliott74}

Nam has noticed that, when evaluating Eq.~\eqref{eq:responsefunction}
within the Eliashberg theory, it is advantageous to replace the
Eliashberg functions $Z(\omega)$ and $\Delta(\omega)$ with three
complex functions: energy function $\epsilon(\omega)$, density of
states $n(\omega)$, and density of pairs $p(\omega)$ defined by
\begin{eqnarray*}
\epsilon(\omega)&=&Z(\omega)\sqrt{\omega^2-\Delta^2(\omega)},
\\
n(\omega)&=&\frac{\omega}{\sqrt{\omega^2-\Delta^2(\omega)}},
\\
p(\omega)&=&\frac{\Delta(\omega)}{\sqrt{\omega^2-\Delta^2(\omega)}}.
\end{eqnarray*}
Note that $n^2(\omega)-p^2(\omega)=1$ and there is some redundancy in
this formulation. Taking into account that the square root is to be
taken so that its imaginary part is positive, one can check readily
the symmetry properties
$$
\epsilon(-\omega)=-\epsilon^\ast(\omega),
\quad
n(-\omega)=n^\ast(\omega),
\quad
p(-\omega)=-p^\ast(\omega).
$$

In terms of the functions $\epsilon(\omega)$, $n(\omega)$, and
$p(\omega)$, the Nambu-Gor'kov Green's function of the Eliashberg
superconductors can be written as
\begin{eqnarray}
\hat{G}({\bf k},\omega) = \frac{1}{2}
\bigg[\frac{n(\omega)\tau_0+p(\omega)\tau_1+\tau_3}
{\epsilon(\omega)-\varepsilon_{\bf k}}
\nonumber
\\
+\frac{n(\omega)\tau_0+p(\omega)\tau_1-\tau_3}
{\epsilon(\omega)+\varepsilon_{\bf k}}\bigg].
\label{eq:green_nam}
\end{eqnarray}

Making use of Eq.~\eqref{eq:green_nam}, Nam has succeeded to express
the optical conductivity in terms of the functions $n(\omega)$,
$p(\omega)$, and $\epsilon(\omega)$. His result can be written in the
following, slightly modified form:
\begin{equation}
\sigma(\omega) = 
\pi D \delta(\omega)+\sigma_{\rm reg}(\omega).
\label{eq:conductivity}
\end{equation}
The first term describes the singular contribution of the condensate
to frequency-dependent conductivity. Its magnitude can be written in
terms of the superfluid density $n_s$ as $D=n_se^2/m$ and is given by
\begin{eqnarray}
\frac{D}{D_0}= \frac{n_s}{n}=
-\int_0^\infty d\nu \tanh\left(\frac{\nu}{2T}\right){\rm Re}
\left[\frac{p^2(\nu)}{\epsilon(\nu)}\right].
\label{eq:singular}
\end{eqnarray}
In the non-superconducting state obviously $D=n_s=0$.

The second term in Eq.~\eqref{eq:conductivity} represents the
non-singular part of the conductivity and is given by\cite{Carbotte05}
\begin{eqnarray}
\sigma_{\rm reg}(\omega)=\frac{iD_0}{\omega}\int_{-\infty}^{\infty}d\nu
\tanh\left(\frac{\nu}{2T}\right) H(\nu+\omega,\nu),
\label{eq:regular}
\end{eqnarray}
where we have introduced an auxiliary complex function
\begin{eqnarray}
H\big(x, y\big)&=& 
\frac{1 + n(x)n^*(y) + p(x)p^*(y)}
{2\left[\epsilon^*(y) - \epsilon(x)\right]},
\nonumber
\\
&+&
\frac{1 - n(x)n(y) - p(x)p(y)}
{2\left[\epsilon(y)+\epsilon(x)\right]}.
\label{eq:function_h}
\end{eqnarray}
The formulae
Eqs.~(\ref{eq:conductivity},\ref{eq:singular},\ref{eq:regular},\ref{eq:function_h})
are valid for any Eliashberg superconductor. It should be noted that
for $\omega\rightarrow 0$, Eq.~\eqref{eq:regular} predicts that
$\sigma_{\rm reg}''(\omega)\approx D/\omega$, which describes the
inductive response of the condensate.

\begin{figure*}[t]
\centering
\includegraphics[width = 15cm]{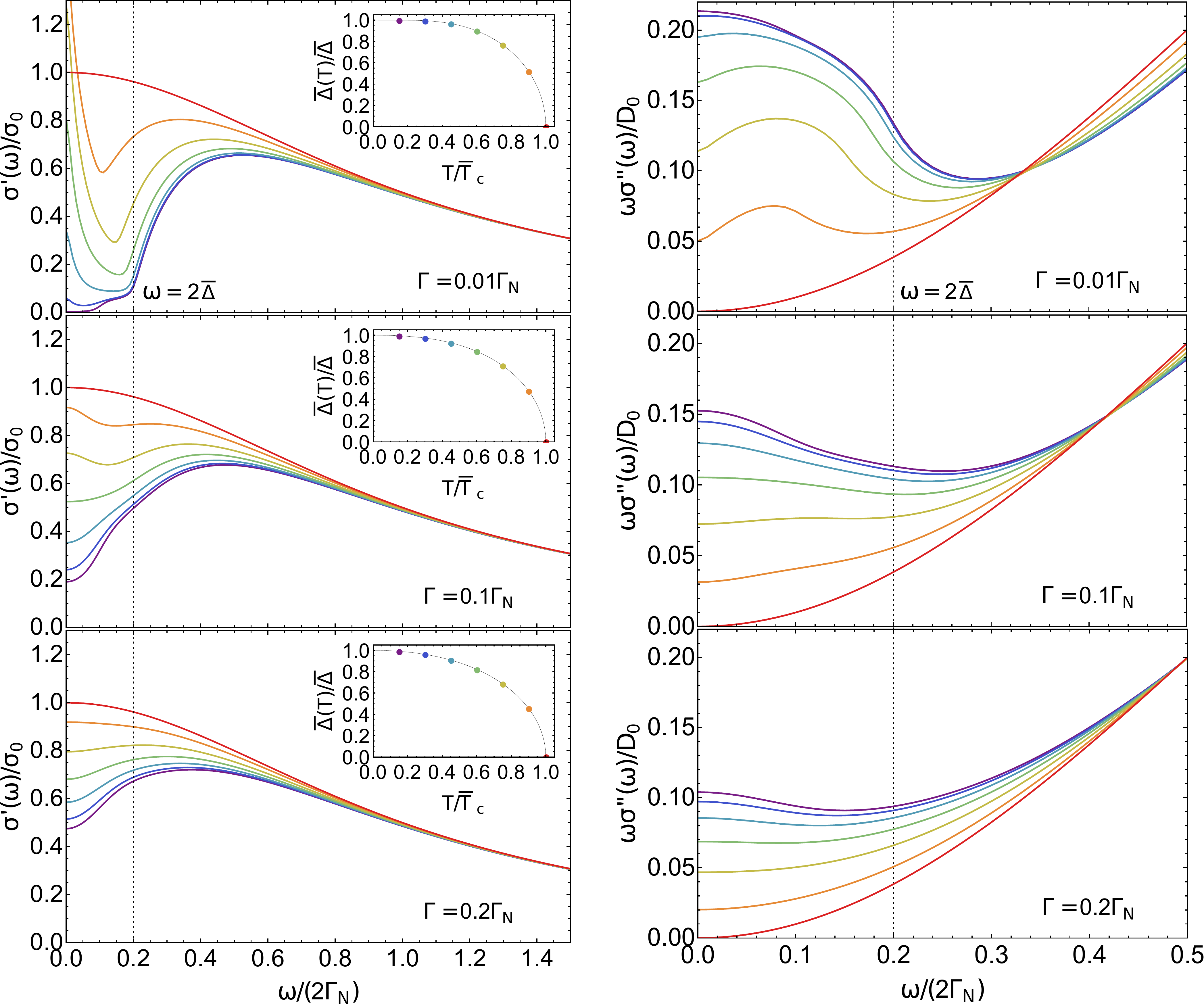}
\caption{Frequency dependence of the real and imaginary parts of the
  conductivity Eq.~\eqref{eq:regular} for the Dynes superconductors
  with fixed total scattering rate $\Gamma_n=5\overline{\Delta}$ and
  varying pair-breaking rates $\Gamma$.  The temperature dependence of
  the gap, $\overline{\Delta}(T)$, was calculated using
  Eq.~\eqref{eq:self-consistent}. The studied temperatures and the
  corresponding $\Gamma$-dependent values of $\overline{\Delta}(T)$
  are presented in the insets. Note that, in all panels, the
  highest-temperature curves represent the normal state. The energy
  scale $2\overline{\Delta}$ is shown by the dotted lines. Let us
  point out that the values of $\omega\sigma''(\omega)/D_0$ at
  $\omega\rightarrow 0$ allow for a direct determination of the
  superfluid fraction $n_s/n$.}
\label{fig:sigma_temp}
\end{figure*}

Before concluding this Section it is worth pointing out that Nam's
theory satisfies the conductivity sum rule
\begin{equation}
\int_{0}^{\infty}d\omega \sigma'(\omega) = \frac{\pi}{2}D_0.
\label{eq:cond_sum_rule}
\end{equation}
The proof of Eq.~\eqref{eq:cond_sum_rule} makes use of two
observations following from
Eqs.~(\ref{eq:kubo},\ref{eq:responsefunction}): (i) the function
$\sigma(\omega)$ satisfies the Kramers-Kronig relations, and (ii) the
high-frequency limit of $\sigma''(\omega)$ is $D_0/\omega$.  Writing
down the Kramers-Kronig relation for $\sigma''(\omega)$ at $\omega
\rightarrow \infty$, one recognizes easily that we arrive at
Eq.~\eqref{eq:cond_sum_rule}.

\section{Optical conductivity: Dynes superconductors}
Now we apply the general expressions from the previous Section to the
Dynes superconductors, for which the energy function
$\epsilon(\omega)$ is given by Eq.~\eqref{eq:epsilon}. The gap
function $\Delta(\omega)$ of the Dynes superconductors is also
known:\cite{Herman16}
\begin{equation}
\Delta(\omega)= \frac{\omega}{\omega + i\Gamma}\overline{\Delta},
\label{eq:delta}
\end{equation}
and this implies that
\begin{eqnarray}
n(\omega)=\frac{\omega+i\Gamma}
{\sqrt{(\omega+i\Gamma)^2-\overline{\Delta}^2}},
\quad
p(\omega)=\frac{\overline{\Delta}}
{\sqrt{(\omega+i\Gamma)^2-\overline{\Delta}^2}}.
\label{eq:dynes_np}
\end{eqnarray}
Typical frequency dependence of the real and imaginary parts of
$n(\omega)$, $p(\omega)$, and $\epsilon(\omega)$ of a Dynes
superconductor is shown in the Appendix.
Equations~(\ref{eq:conductivity},\ref{eq:singular},\ref{eq:regular},\ref{eq:function_h})
together with Eqs.~(\ref{eq:epsilon},\ref{eq:dynes_np}) provide a
complete description of the electromagnetic properties of the Dynes
superconductors. Note that a single integration is required to
determine the optical conductivity $\sigma(\omega)$, therefore the
numerical cost is the same as in the widely used Mattis-Bardeen
theory,\cite{Mattis58} which can be viewed as a special limit
($\Gamma=0$ and $\Gamma_n\gg\overline{\Delta}$) of the present
approach.

\begin{figure*}[t]
\includegraphics[width = 14cm]{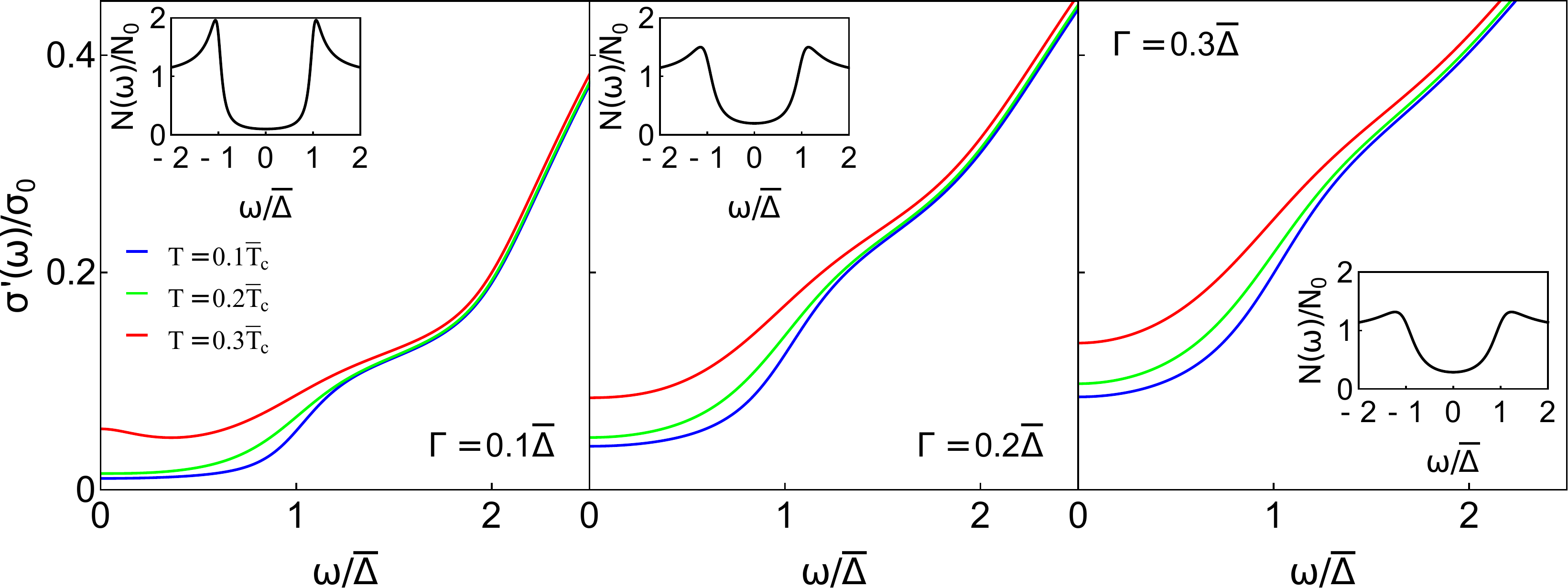}
\caption{Frequency dependence of the dirty-limit conductivity
  $\sigma'_{\rm reg}(\omega)$, Eq.~\eqref{eq:dirty}, at low
  frequencies. For each considered value of the pair-breaking rate
  $\Gamma$ several temperatures are studied, and the corresponding
  Dynes density of states at $T=0$ is shown in the inset.  The
  absorption edge at $\omega=\overline{\Delta}$ is clearly visible for
  $\Gamma\lesssim 0.3\overline{\Delta}$ and $T\lesssim
  0.3\overline{T_c}$.}
\label{fig:edge}
\end{figure*}

As shown in Ref.~\onlinecite{Herman16}, the magnitude of the gap
$\overline{\Delta}$ of a Dynes superconductor is controlled by the
pair-breaking scattering rate $\Gamma$, and in the BCS model with
coupling constant $\lambda$ and cutoff frequency $\Omega$ it can be
found by solving the self-consistent equation
\begin{equation}
\overline{\Delta}(T)=2\lambda\pi T \sum_{\omega_n>0}^\Omega
\frac{\overline{\Delta}(T)}
{\sqrt{(|\omega_n|+\Gamma)^2+\overline{\Delta}^2(T)}},
\label{eq:self-consistent}
\end{equation}
where the sum is taken over the Matsubara frequencies $\omega_n$.
Starting with Eq.~\eqref{eq:self-consistent}, we use the following
notation: $\overline{\Delta}(T)$ denotes the gap at temperature $T$,
whereas the $T=0$ value of the gap will be denoted
simply as $\overline{\Delta}$.  Moreover, the critical temperature of
the dirty system will be called $\overline{T_c}$.

Before discussing the optical conductivity in the superconducting
state, it is useful to start by analyzing the normal state
$\overline{\Delta}=0$. In this case we have
$\epsilon(\omega)=\omega+i\Gamma_n$, $n(\omega)=1$, and $p(\omega)=0$,
and therefore the normal-state optical conductivity of a Dynes
superconductor is given by the simple Drude formula
\begin{equation}
\sigma_n(\omega) = \frac{D_0}{2\Gamma_n - i\omega}.
\label{eq:sigma_normal}
\end{equation}
Note that the relative weight of pair-breaking and pair-conserving
processes is irrelevant in the normal state and therefore
$\sigma_n(\omega)$ depends only on the total scattering rate
$\Gamma_n$.  In what follows we will measure the conductivities in
terms of $\sigma_0=D_0/(2\Gamma_n)$, which is the normal-state
conductivity at $\omega=0$.

Our main goal is to study the change of the optical conductivity
$\sigma(\omega)$ in the superconducting state at fixed $\Gamma_n$ for
varying ratios of pair-conserving and pair-breaking processes.  In
Fig.~\ref{fig:Pedagog} we plot $\sigma^\prime(\omega)$ of a moderately
dirty Dynes superconductor with $\Gamma_n=5\overline{\Delta}$ at
temperature $T=0$ for two values of the pair-breaking scattering rate
$\Gamma$. In absence of pair-breaking, i.e. for $\Gamma=0$, our
results are consistent with the Mattis-Bardeen theory generalized so
as to apply for arbitrary values of $\Gamma_s$:\cite{Zimmermann91} for
$\omega<2\overline{\Delta}$, the conductivity vanishes, whereas for
$\omega\gg\overline{\Delta}$ it approaches the normal-state
value. Note that already very small amount of pair-breaking
$\Gamma=0.02\Gamma_n$ leads to dramatic changes of the low-frequency
conductivity: even at the lowest frequencies, $\sigma^\prime_{\rm
  reg}(\omega)$ is finite, and in addition to the step at
$\omega=2\overline{\Delta}$, optical conductivity also shows an
additional step at $\omega=\overline{\Delta}$.  Both of these results
can be easily understood by inspecting the density of states of the
Dynes superconductor, also shown in Fig.~\ref{fig:Pedagog}: similarly
as in the normal metal, absorption at arbitrarily low frequencies is
possible also in the Dynes superconductor, and the joint density of
states increases at the two observed
steps. Figure~\ref{fig:sigma_temp} shows that the filling in of the
gap by pair-breaking processes quickly grows with $\Gamma$ and that
the effect of pair breaking processes is quite similar to that of
raising the temperature. As a result, the missing weight below the
$\sigma^\prime_n(\omega)$ curve diminishes, resulting (due to the
Ferrell-Glover-Tinkham sum rule\cite{Ferrell58}) in quickly decreasing
superfluid fraction $n_s/n$. 

{\it Dirty limit.} The general formula Eq.~\eqref{eq:regular} is
somewhat difficult to interpret, but in the dirty limit, when
$\Gamma_s\gg\overline{\Delta},\Gamma$, it can be simplified
considerably. In fact, if we restrict ourselves to frequencies
$\omega\ll \Gamma_s$, the function $\epsilon(\omega)$ reduces to
$\epsilon(\omega)\approx i\Gamma_n$, and in this case
Eq.~\eqref{eq:regular} reduces to the physically much more transparent
formulas
\begin{eqnarray}
\sigma'_{\rm reg}(\omega)&=&\frac{\sigma_0}{\omega}
\int_{-\infty}^\infty d\nu \left[f(\nu)-f(\nu+\omega)\right]
\nonumber
\\
&\times&
\left[n'(\nu)n'(\nu+\omega)+p'(\nu)p'(\nu+\omega)\right],
\nonumber
\\
\sigma''_{\rm reg}(\omega)&=&-\frac{\sigma_0}{\omega}
\int_{-\infty}^\infty d\nu \left[1-2f(\nu)\right]
\nonumber
\\
&\times&
\left[n'(\nu)n''(\nu+\omega)+p'(\nu)p''(\nu+\omega)\right],
\label{eq:dirty}
\end{eqnarray}
where $f(\nu)$ is the Fermi-Dirac distribution function.  Similar
expressions have been guessed without proof also in
Ref.~\onlinecite{Zemlicka15}. The real part $\sigma'_{\rm
  reg}(\omega)$ may be interpreted as a sum of contributions from the
single-particle and Cooper-pair channels, both of which absorb the
radiation in a semiconductor-like fashion. Note that, if one sets the
pair-breaking rate $\Gamma$ to zero in Eqs.~\eqref{eq:dirty}, one
recovers a very compact form of the Mattis-Bardeen theory. It is also
worth pointing out that the optical conductivity $\sigma_{\rm
  reg}(\omega)$ of course does depend, via $\sigma_0$, on the
normal-state scattering rate $\Gamma_n$.

\begin{figure}[t]
\includegraphics[width = 7cm]{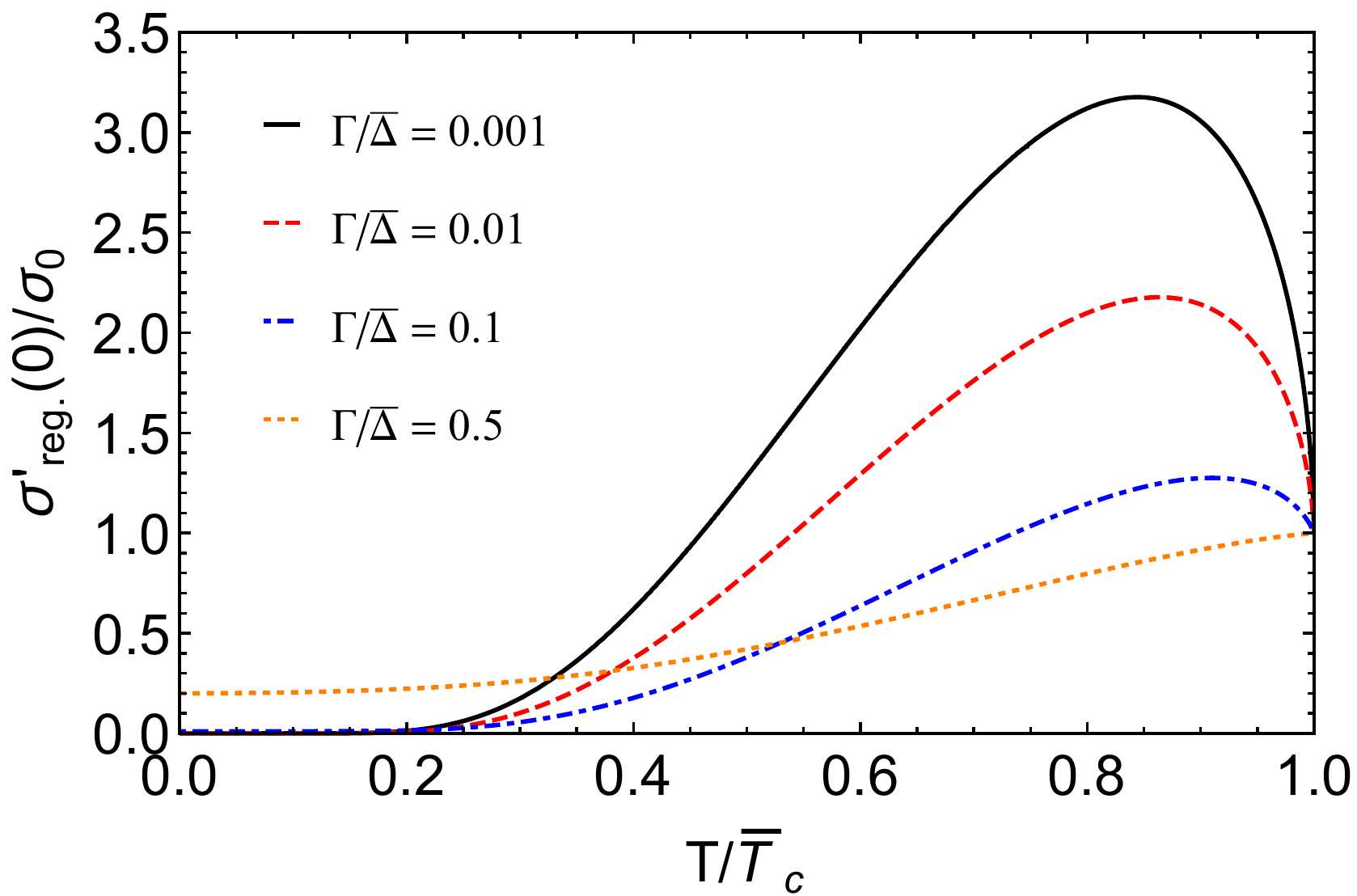}
\caption{Temperature dependence of the dirty-limit conductivity
  $\sigma'_{\rm reg}(0)$, Eq.~\eqref{eq:coherence}, for several values
  of the pair-breaking scattering rate $\Gamma$.}
\label{fig:coherence}
\end{figure}

An observation of the absorption edge at $\omega=\overline{\Delta}$
would provide a smoking gun for the concept of the Dynes
superconductors. In order to identify the conditions under which this
feature can be observable, in Fig.~\ref{fig:edge} we plot
$\sigma^\prime_{\rm reg}(\omega)$ in the experimentally relevant dirty
limit. One observes that the absorption edge at
$\omega=\overline{\Delta}$ is clearly visible for $\Gamma\lesssim
0.3\overline{\Delta}$ and $T\lesssim 0.3\overline{T_c}$, thus
rendering the concept of the Dynes superconductors experimentally
falsifiable by optical means.

\begin{figure}[t]
\includegraphics[width = 7cm]{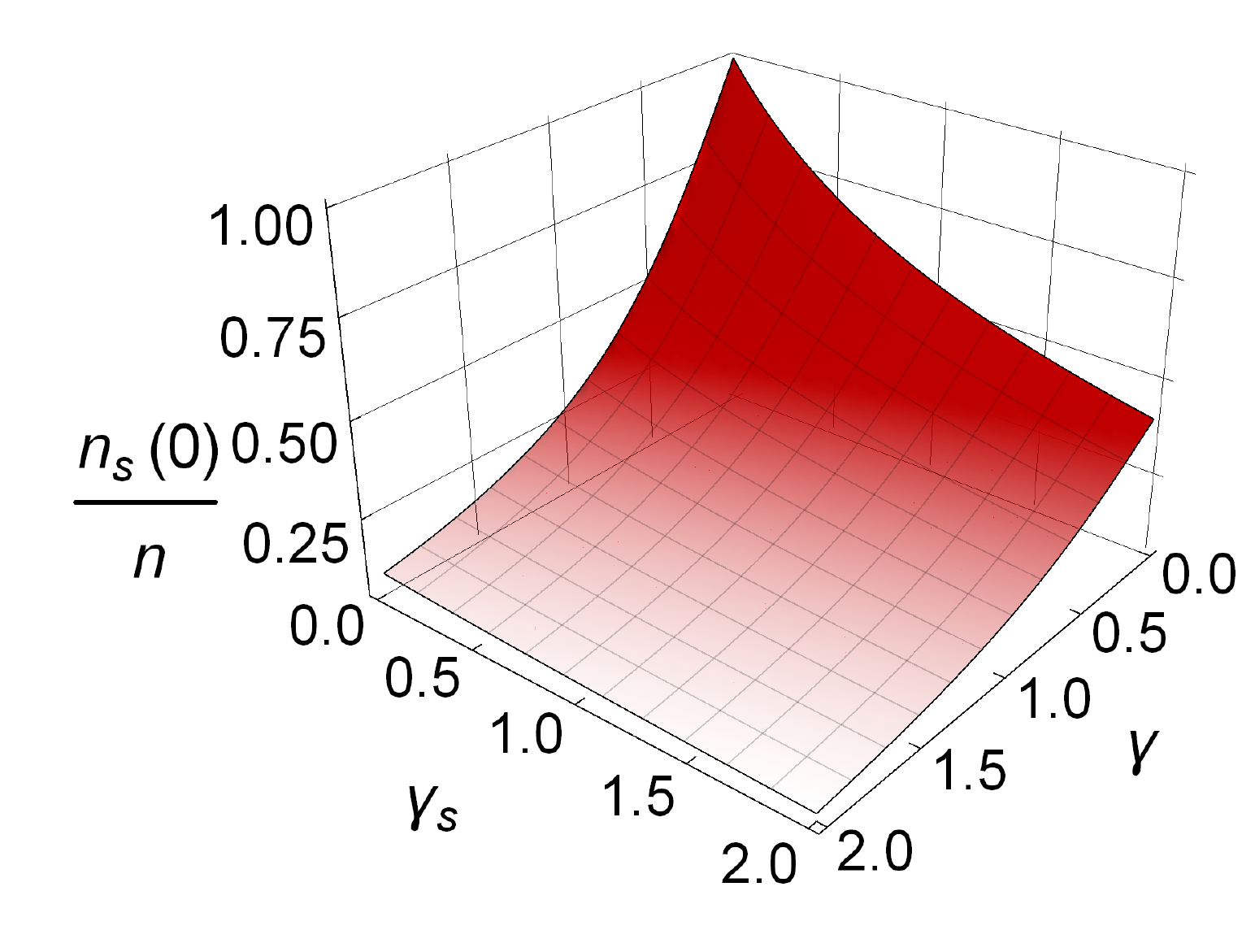}
\caption{Superfluid fraction $n_s/n$ at $T=0$ as a function of the
  scattering rates $\gamma$ and $\gamma_s$ according to
  Eq.~\eqref{eq:nslnDT0}.}
\label{fig:nsln}
\end{figure}

{\it Coherence peak.}  It is well known\cite{Marsiglio08} that in
absence of pair breaking, the low-frequency conductivity $\sigma'_{\rm
  reg}(\omega\rightarrow 0)$ diverges for all temperatures $T<T_c$. In
this paragraph we will calculate $\sigma'_{\rm reg}(0)$ for a Dynes
superconductor, assuming that the dirty limit applies. We will show
that $\sigma'_{\rm reg}(0)$ stays finite, and that, as a function of
temperature, it exhibits the well-known coherence peak with a
magnitude controlled by $\Gamma$. To this end, let us make use of
Eq.~\eqref{eq:dirty} which implies that
\begin{equation}
\frac{\sigma'_{\rm reg}(0)}{\sigma_0}=
\int_0^\infty \frac{dx}{2T\cosh^2(x/2T)}\left[n'(x)^2+p'(x)^2\right].
\label{eq:coherence}
\end{equation}
Results of the numerical evaluation of Eq.~\eqref{eq:coherence} are
presented in Fig.~\ref{fig:coherence}. Note that, for sufficiently
large pair-breaking rates $\Gamma$, the temperature dependence of
$\sigma'_{\rm reg}(0)$ is monotonic, without any coherence peaks.
However, for sufficiently small $\Gamma$, the conductivity
$\sigma'_{\rm reg}(0)$ immediately below $T_c$ grows with decreasing
temperature. In the vicinity of $T_c$, where $\overline{\Delta}$ is
small, the growth is controlled by $\Gamma$, and for $\Gamma\ll T_c$
it can be shown that
\begin{equation}
\frac{\sigma'_{\rm reg}(0)}{\sigma_0}=
1+\frac{\pi}{8}\frac{\overline{\Delta}^2(T)}{\overline{T_c}\Gamma}.
\label{eq:coherence_high}
\end{equation}
When the temperature is decreased further, the thermal factor in
Eq.~\eqref{eq:coherence} starts to dominate and $\sigma'_{\rm reg}(0)$
decreases, until ultimately in the low-$T$ region it saturates at
\begin{equation}
\frac{\sigma'_{\rm reg}(0)}{\sigma_0}
=\frac{\Gamma^2}{\Gamma^2+\overline{\Delta}^2}.
\label{eq:coherence_low}
\end{equation}
The formula Eq.~\eqref{eq:coherence_low} is valid in the dirty limit
for any ratio between $\Gamma$ and $\overline{\Delta}$. Note that the
results Eqs.~(\ref{eq:coherence_high},\ref{eq:coherence_low}) can be
used for a direct determination of the pair-breaking rate $\Gamma$ by
microwave measurements.

{\it Superfluid fraction.} The superfluid fraction $n_s/n$ of a Dynes
superconductor can be determined from Eq.~\eqref{eq:singular}.  In the
limit $T=0$, the integral can be taken analytically and, introducing
dimensionless scattering rates $\gamma = \Gamma/{\overline{\Delta}}$
and $\gamma_s = \Gamma_s/{\overline{\Delta}}$, the result can be
written as
\begin{widetext}
\begin{eqnarray}
\frac{n_s(0)}{n} = \begin{cases}  
\frac{1}{\gamma_s}\Bigg[
\arctan(1/\gamma)-\frac{1}{\sqrt{1-\gamma_s^2}}
\bigg(\arccos\gamma_s
+\arctan\frac{\sqrt{1-\gamma_s^2}}{\gamma}
-\arctan\frac{\sqrt{1-\gamma_s^2}\sqrt{\gamma^2+1}}
{\gamma\gamma_s}
\bigg)
\Bigg] &
\text{if } \gamma_s< 1, 
\\
\frac{1}{\gamma_s}\Bigg[
\arctan(1/\gamma)-\frac{1}{\sqrt{\gamma_s^2-1}}
\ln\frac{\big(\gamma_s+\sqrt{\gamma_s^2-1}\big)
\big(\gamma+\sqrt{\gamma_s^2-1}\big)}
{\gamma\gamma_s+\sqrt{\gamma_s^2-1}\sqrt{\gamma^2+1}}
\Bigg] & 
\text{if } \gamma_s\geq 1.     
        \end{cases}
\label{eq:nslnDT0}
\end{eqnarray}
\end{widetext}
The formula Eq.~\eqref{eq:nslnDT0} is somewhat cumbersome. In order to
illustrate its meaning, in Fig.~\ref{fig:nsln} we present a 3D plot of
$n_s/n$ at $T=0$ as a function of $\gamma$ and $\gamma_s$.  Notice
that both types of scattering processes diminish the superfluid
fraction, but (as expected) the pair-breaking impurities are much more
effective in doing so. We have also checked that in absence of
pair-breaking processes, i.e. for $\gamma=0$, Eq.~\eqref{eq:nslnDT0}
reduces to the previously published
results.\cite{Nam67a,Marsiglio08,Berlinsky93}

For the sake of completeness let us also point out that, instead of
using Eq.~\eqref{eq:singular}, at finite temperatures the superfluid
fraction of a Dynes superconductor can be more conveniently calculated
from the Matsubara sum
\begin{equation}
\frac{n_s(T)}{n}=2\pi T\sum_{\omega_n>0}\frac{\overline{\Delta}^2}
{\Omega_n^2\left(\Omega_n+\Gamma_s\right)},
\label{eq:fraction_finite_T}
\end{equation}
where $\Omega_n=\sqrt{(|\omega_n|+\Gamma)^2+\overline{\Delta}^2}$.
Note that Eq.~\eqref{eq:fraction_finite_T} can be obtained from the
well-known result for superconductors in which only pair-conserving
scattering is present,\cite{Nam67a,Abrikosov63} provided one takes the
pair-breaking processes into account by the minimal substitution
$|\omega_n|\rightarrow |\omega_n|+\Gamma$, which was mentioned in the
Introduction.

\begin{figure}[t]
\centering
\includegraphics[width = 8.5cm]{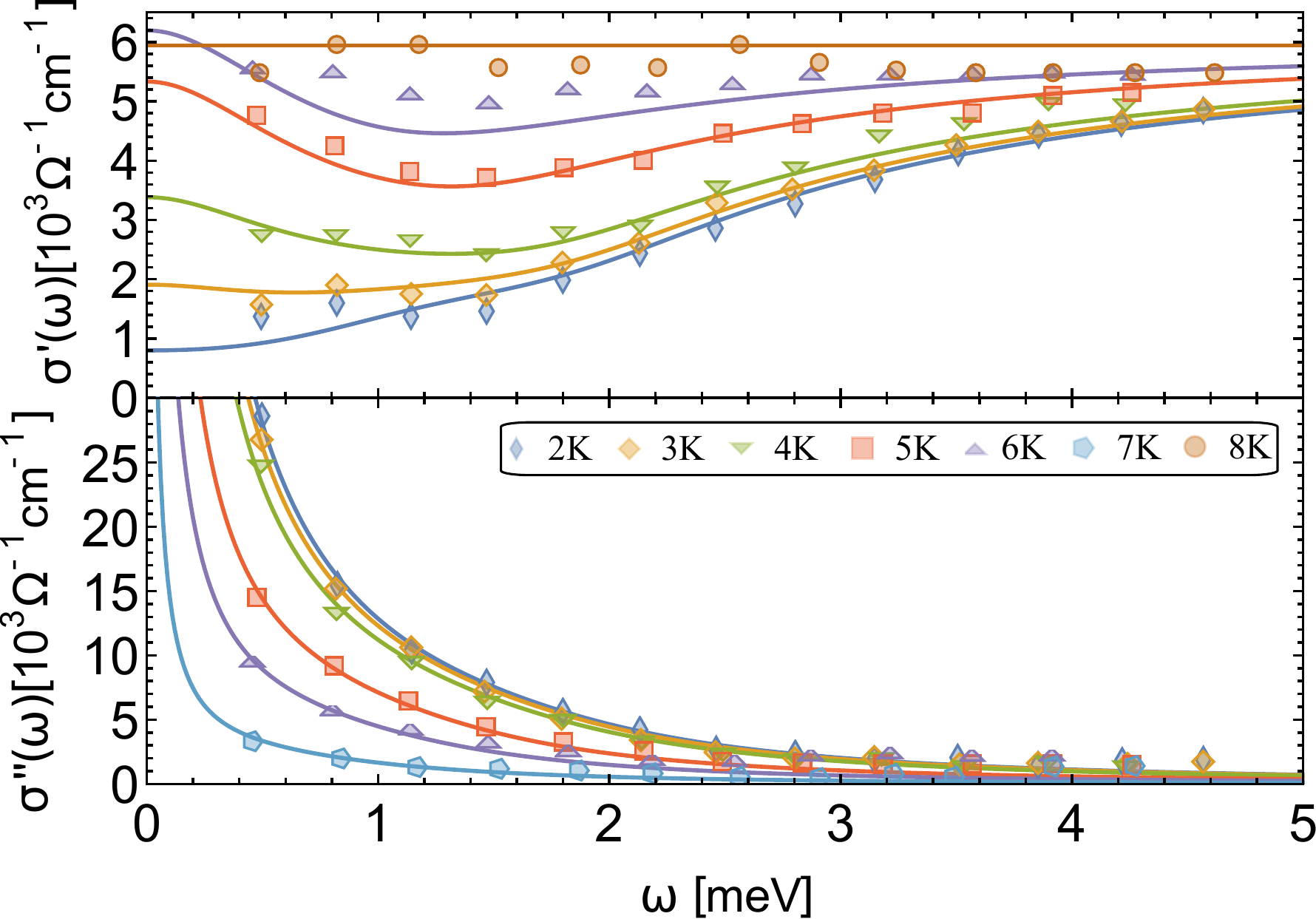}
\caption{Real and imaginary parts of the optical conductivity for a
  $15.1$~nm thick MoN sample reported in
  Ref.~\onlinecite{Simmendinger16}. Fits of the data to the prediction
  for a dirty Dynes superconductor, Eqs.~\eqref{eq:dirty}, are also
  shown. At $T=2$~K, the fits are three-parametric. For all curves at
  higher temperatures there is only one fitting parameter, namely
  $\overline{\Delta}(T)$.}
\label{fig:MoN_sigma}
\end{figure}

\section{Application to Molybdenum Nitride}
Motivated by interest in the physics of the superconductor-insulator
transition,\cite{Lin15} strongly disordered superconducting thin films
have recently become the subject of intensive research. Since in-gap
states are frequently observed in such
systems\cite{Noat13,Szabo16,Lin15,Cheng16} and since their tunneling
density of states can be often described by the Dynes formula
Eq.~\eqref{eq:dynes},\cite{Noat13,Szabo16} the theory developed in
this paper should be directly applicable precisely to such systems. It
is also fortunate that, due to progress in terahertz spectroscopy,
high-quality optical data on strongly disordered superconductors have
recently become available.\cite{Cheng16,Simmendinger16}

As an example, in this paper we have chosen to discuss the terahertz
spectra obtained by Simmendinger {\it et al.} on molybdenum nitride
(MoN) thin films in Ref.~\onlinecite{Simmendinger16}. The authors have
studied a whole series of films with different thickness, but most
details were presented for the 15.1~nm thick film, and we will test
the predictions of our theory against this sample.

Simmendinger {\it et al.} start their analysis by noting that although
their films are definitely deeply in the dirty limit, the
Mattis-Bardeen theory can not fit their data. The reason is that even
at the lowest temperatures, large in-gap absorption has been observed.
In Ref.~\onlinecite{Simmendinger16} this finding has been explained by
the existence of a hypothetical parallel normal-conducting channel
inside the sample. While this might be possible - although one would
have to explain the absence of an internal proximity coupling between
the normal channel and the superconducting channel - we consider here
an alternative explanation, namely that MoN is a Dynes superconductor.

We proceed as follows: since the studied sample is obviously in the
extremely dirty limit, we can use Eqs.~\eqref{eq:dirty} to fit the
observed frequency dependence of the real and imaginary parts of
conductivity.  At the lowest experimental temperature $T=2$~K the
least-squares fitting procedure is robust and we can safely fix the
three free parameters entering Eqs.~\eqref{eq:dirty}, namely the
overall scale $\sigma_0$, the superconducting gap $\overline{\Delta}$,
and the pair-breaking scattering rate $\Gamma$.  The resulting fit is
shown in Fig.~\ref{fig:MoN_sigma}; the agreement between theory and
experiment is seen to be reasonable.

\begin{figure}[t]
\centering
\includegraphics[width = 7cm]{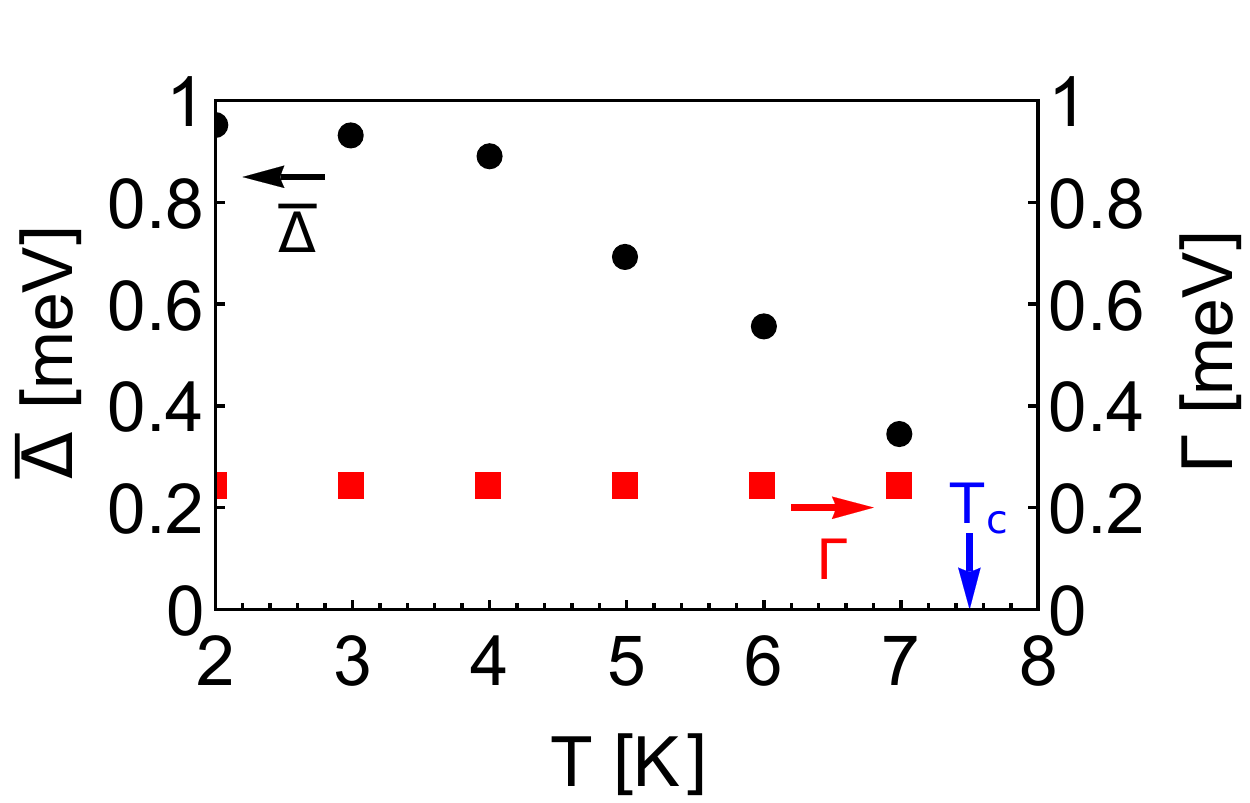}
\caption{Temperature dependence of the gap parameter
  $\overline{\Delta}(T)$ obtained from the fits shown in
  Fig.~\ref{fig:MoN_sigma}. Also shown is the value of $\Gamma$ which
  has been determined at the lowest temperature $T=2$~K; at higher
  temperatures its value was kept fixed.}
\label{fig:MoN_delta}
\end{figure}

Next we assume that both $\sigma_0$ and $\Gamma$ do not depend on
temperature, and using their values which have been determined at
$T=2$~K, we fit the higher-temperature conductivity data by a single
parameter: the superconducting gap $\overline{\Delta}(T)$. The
resulting temperature dependence of the superconducting gap
$\overline{\Delta}(T)$ exhibits the expected order parameter-like
shape, see Fig.~\ref{fig:MoN_delta}, and the corresponding fits of the
optical conductivity are shown in Fig.~\ref{fig:MoN_sigma}. The
agreement between experiment and theory is satisfactory, especially if
we realize that the fits of two frequency-dependent functions
$\sigma'_{\rm reg}(\omega)$ and $\sigma''_{\rm reg}(\omega)$ are done
using a single free parameter $\overline{\Delta}(T)$ at every
temperature $T>2$~K.

Note that our estimate of the pair-breaking rate of the 15.1~nm thick
MoN film, $\Gamma\approx 0.2$~meV, is of the same order of magnitude
as $\Gamma\approx 0.1$~meV directly measured by tunneling in a similar
10~nm thick MoC film.\cite{Szabo16} It is worth pointing out that such
values are reasonable. In fact, according to our interpretation,
$\Gamma$ is a typical exchange field inside the superconducting
film. If we assume that the magnetic impurities are located in the
vicinity of the film/substrate interface with area density $n_\Box$,
and if we characterize the impurities by exchange field $J$ decaying
to zero on the length scale $a$ (of atomic dimensions), then we can
make the following very rough estimate: $\Gamma\approx J a^3
n_\Box/d$. Let us take the typical values $J=1$~eV, $a=0.3$~nm, and
for the film thickness $d=15.1$~nm. If we assume that $n_\Box=x/a^2$
(where $x$ is the fraction of atomic positions occupied by magnetic
impurities), then we obtain that $\Gamma\approx 0.2$~meV corresponds
to $x=0.01$, which looks quite reasonable. Larger values of $a$ lead
to even smaller concentrations of the magnetic impurities.

\section{Conclusions}
Based on Nam's description of the electromagnetic properties of
superconductors,\cite{Nam67a} in this paper we have presented a
comprehensive set of predictions for the optical conductivity of the
recently identified Dynes superconductors.\cite{Herman16,Herman17} In
particular, we have shown that two metals with the same optical
response in the normal state and equal superconducting gaps
$\overline{\Delta}$ may exhibit very different superconducting
responses, with the shape of the latter depending on the ratio of the
pair-breaking and pair-conserving scattering rates $\Gamma$ and
$\Gamma_s$, see Fig.~\ref{fig:sigma_temp}.

The most characteristic optical fingerprint of a Dynes superconductor
is the presence (at low temperatures and for physically reasonable
pair-breaking rates) of an additional absorption edge in
$\sigma'(\omega)$ at $\omega=\overline{\Delta}$, which exists in
addition to the conventional absorption edge at
$\omega=2\overline{\Delta}$.  Another property which is unique to a
Dynes superconductor is that the dissipative component of the
low-frequency conductivity $\sigma'_{\rm reg}(\omega\rightarrow 0)$
stays finite down to temperature $T=0$. Both of these anomalies are in
fact a simple consequence of the fact that the Dynes superconductors
are gapless, see Fig.~\ref{fig:Pedagog}.

Furthermore we have shown that the pair-breaking scattering rate
$\Gamma$ can be straightforwardly determined from microwave
measurements, either from the slope of the coherence peak,
Eq.~\eqref{eq:coherence_high}, or from the low-frequency conductivity
$\sigma'_{\rm reg}(\omega\rightarrow 0)$ in the limit of low
temperatures, Eq.~\eqref{eq:coherence_low}, thus enabling comparison
with the $\Gamma$ values obtained from the tunneling spectroscopy.

Practical formulae have been derived for the superfluid fraction
$n_s/n$ (or, equivalently, superfluid stiffness) of the Dynes
superconductors. In the zero-temperature limit, we have found an
explicit algebraic expression, Eq.~\eqref{eq:nslnDT0}, showing how
$n_s(0)/n$ depends on the scattering rates $\Gamma$ and $\Gamma_s$.
At finite temperatures, the formula Eq.~\eqref{eq:fraction_finite_T},
which is suitable for an efficient numerical evaluation of $n_s(T)/n$,
has been derived.

\begin{figure}[t]
\includegraphics[width = 7 cm]{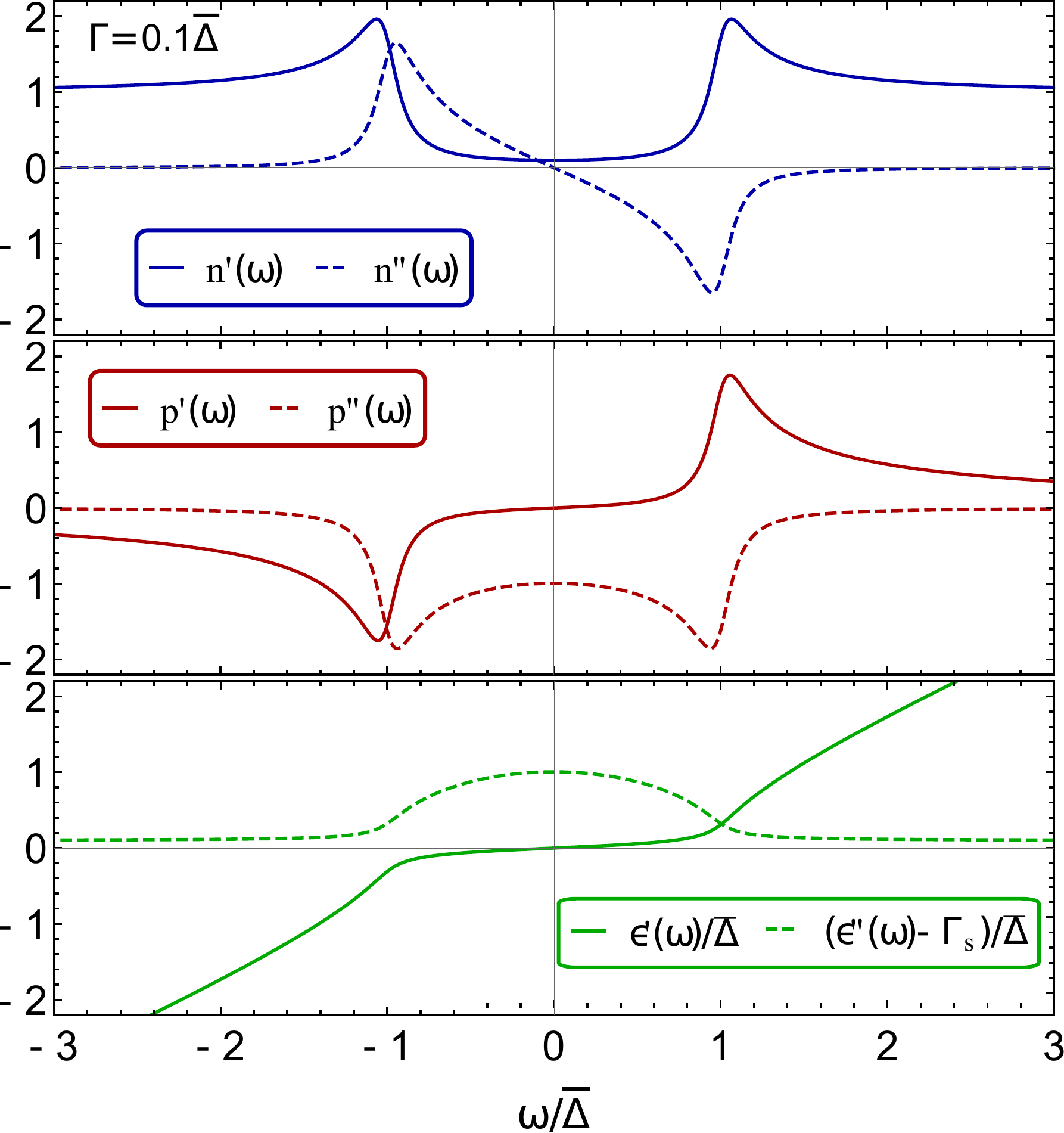}
\caption{The real and imaginary parts of the functions $n(\omega)$,
  $p(\omega)$, and $\epsilon(\omega)$ describing a Dynes
  superconductor with the pair-breaking rate
  $\Gamma=0.1\overline{\Delta}$.}
\label{fig:appendix}
\end{figure}

Strongly disordered thin superconducting films seem to be the best
candidate where the Dynes phenomenology may be
observable.\cite{Szabo16,Noat13} This is probably caused by the
presence of magnetic impurities at the interface between the film and
the substrate, which may be present even in otherwise very clean
films.\cite{Casalbuoni05,Junginger17} In this paper we have shown that
the apparently anomalous optical data for MoN thin
films\cite{Simmendinger16} can be reasonably fitted by the Dynes
optics, see Figs.~\ref{fig:MoN_sigma},\ref{fig:MoN_delta}. This result
lends further support to the identification of strongly disordered
thin superconducting films as potential Dynes superconductors.

From the methodological point of view, we would like to point out that
Eqs.~(\ref{eq:conductivity},\ref{eq:singular},\ref{eq:regular},\ref{eq:function_h})
together with Eqs.~(\ref{eq:epsilon},\ref{eq:dynes_np}) provide a
complete description of the electromagnetic properties of the Dynes
superconductors.  Their numerical evaluation is equally costly as that
which makes use of the generalized Mattis-Bardeen
formula,\cite{Zimmermann91} but, unlike the latter, allows also for
pair-breaking processes. As regards the formal properties of our
results, the optical conductivity Eq.~\eqref{eq:conductivity} has the
correct analytic properties and high-frequency asymptotics, and
therefore it satisfies also the conductivity sum rule
Eq.~\eqref{eq:cond_sum_rule}. Moreover, by scanning a wide range of
parameters $\overline{\Delta}$, $\Gamma_s$, and $\Gamma$, we have
checked that $\sigma'(\omega)$ for a Dynes superconductor is positive
definite (as it should be), although we were not able to prove it.
This means that our theory for optics of the Dynes superconductors
satisfies the same set of constraints which is obeyed by the simple
Drude formula. Since the latter is known to be a good starting point
when analyzing the optics of normal metals, we are convinced that the
present theory might play an analogous role in the superconducting
state.

\begin{acknowledgments}
This work was supported by the Slovak Research and Development Agency
under contracts No.~APVV-0605-14 and No.~APVV-15-0496, and by the
Agency VEGA under contract No.~1/0904/15.
\end{acknowledgments}

\appendix*
\section{The functions $n(\omega)$, $p(\omega)$, and $\epsilon(\omega)$}
For readers' convenience, in Fig.~\ref{fig:appendix} we plot the
functions $n(\omega)=n^\prime(\omega)+i n^{\prime\prime}(\omega)$ and
$p(\omega)=p^\prime(\omega)+i p^{\prime\prime}(\omega)$ for a Dynes
superconductor, defined by Eq.~(\ref{eq:dynes_np}).  When frequency is
measured in units of $\overline{\Delta}$, these functions depend on a
single parameter: the pair-breaking rate $\Gamma$. Also shown in
Fig.~\ref{fig:appendix} are the real and imaginary parts of the
function $\epsilon(\omega)$ defined by Eq.~(\ref{eq:epsilon}).

\end{document}